# Analyzing Performance Properties Collected by the PerSyst Scalable HPC Monitoring Tool

David Brayford, Christoph Bernau, Wolfram Hesse, , Carla Guillen

Leibniz-Rechenzentrum der Bayerischen Akademie der Wissenschaften
Munich, Germany
brayford@lrz.de, bernau@lrz.de, hesse@lrz.de, guillen@lrz.de

*Abstract*—The ability to understand how a scientific application is executed on a large HPC system is of great importance in allocating resources within the HPC data center. In this paper, we describe how we used system performance data to identify: execution patterns, possible code optimizations and improvements to the system monitoring. We also identify candidates for employing machine learning techniques to predict the performance of "similar" scientific codes.

## I. Introduction

HPC data centers routinely employ system monitoring tools, to ensure their systems remain stable, operational and that the resources are being used in accordance with the agreed upon contract and the data centers policies. So, it is vital that the data center is able to determine whether the system is being used as intended. However, in practice it is almost impossible to determine exactly what is being executed on the HPC system, because most system monitoring tools do not analyze the application executing on the system. The tools that do, require the code to be instrumented as well as having significant performance overheads, which means that they are not suitable for use on large scale production HPC systems.

PerSyst was developed at LRZ [1] to be a scalable, non-instrumented and low overhead application monitoring tool and has been deployed on the different HPC systems at LRZ since 2010 to collect information from applications that have run on these HPC systems.

In this paper, we perform statistical analysis of the data collected by the PerSyst tool over the past two years. This information is then used to identify how the scientific applications are executed on a large HPC system to infer how the application is being used. This includes, software development, optimizations, scaling tests and production runs. This information can also be used to determine the effectiveness of the PerSyst monitoring tool and if the data collection techniques require modifications. Finally, can we use the information acquired to determine whether or not the application has been optimized or not [2, 3]. For example, we are able to determine if a linear algebra based application is utilizing the SIMD instructions on the CPU.

## II. PerSyst tool

There are several challenges associated with efficient monitoring of large scale production HPC systems and the most important issue is scalability [4-6]. Other issues include the requirement of automatic and online detection of applications containing bottlenecks, with low resource overheads [7, 8]. However, there is a lack of tools that provide on-line analysis without significant overheads and without instrumentation of user codes for the collection of data on a system wide basis.

The PerSyst monitoring tool provides a method of addressing the challenges faced by system monitoring of large production HPC machines and the analysis of important scientific applications executed on SuperMUC with negligible overhead.

PerSyst has been developed as distributed software with a tree agent hierarchy. The three types of agents are the synchronization agent, or SyncAgent; the collector agent; and the PerSyst agent, as shown in figure 1. The main functionalities of the SyncAgent are to synchronize measurement, the collector agents collect the performance data, and the PerSyst agents perform the measurements. All of the agents comprise the transport system for analyzed performance data.

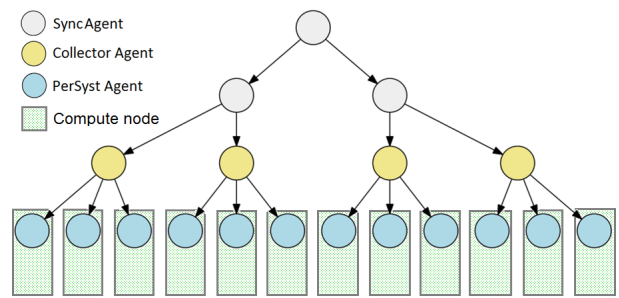

Fig. 1. A graphical represntation of the agent hierarchy

The software is designed to have one layer of PerSyst agents, one layer of collector agents which manage the PerSyst agents, and there is at least one SyncAgent as the frontend. The term middle layers will be used in this chapter for all the agents which are between the front end and the PerSyst agents. For systems in the range of hundreds of thousands of cores, it is necessary to include more layers of SyncAgents in the middle layers of the hierarchy as illustrated in figure 1.

The main functionality of the PerSyst tool is to analyze, filter, collect, and aggregate performance data. The aggregation uses a fixed number of quantiles for all cores in a job. It is known that aggregation of subsets of quantiles is only possible in special cases and can only be done using estimations of the accumulated distributed frequency of each subset. Thus, two types of aggregations are used: estimation of quantiles or exact calculation. The SyncAgents only perform estimation of quantiles, while the collector agents and PerSyst agents perform exact calculation of quantiles.

Data acquired by the monitoring is stored for the entire life-span of a system so that the development of one application can be compared. This enables the observation of the evolution of a code with respect to performance when run with different parameters, number of cores, and data input sets. A system like SuperMUC with 140,000 cores, which collects 40 metrics, will generate approximately 23MiB in one time point. If metrics are collected once per second, the total data collected daily will be approximately 1.96TB. To overcome the problem of data growth, the frequency of data collection was set at 10 minutes for SuperMUC.

*A. Related Work*

PerSyst [1, 9] is derived from the profiling based tool Periscope [10]. Periscope is a scalable tool for analyzing application performance. It enables a distributed online search for performance properties based on hardware counters as well as MPI properties. Periscope is not a monitoring system, but some of the functionality of property processing in PerSyst Monitoring is similar to Periscope. A clear difference between them is that the latter one utilizes an instrumented code. It provides the possibility of defining a user region within the code [11]. There are also potential look-up regions within the instrumented section of the code which are automatically detected. These regions, also called phase regions, are analyzed and refined when bottlenecks are detected in order to delimit the section of the code which presents a problem. In a high performance environment, there is no inherent instrumentation of the applications running, therefore the users shouldn't be requested to instrument their codes as this causes overhead. The analysis and monitoring is done without localizing the code regions of potential stalls and weaknesses at each application. PerSyst Monitoring provides analysis at intervals at a system level, collectible per job, in order to have a first outlook on which applications need tuning. The closer examination at an application level would be the next step to our monitoring and analysis system by means of instrumentation and the use of a performance tool like Periscope. Other system monitoring solutions have been developed. The system most related with our approach made by Nataraj et al. [5] for application monitoring has also a similar back end system monitoring by means of Supermon. This component, Supermon, has a scalability of up to 2048 [12] and handles client requests to extract performance information from the kernel. While PerSyst Monitoring is based on a cyclic system-wide measurement rationale, Supermon is driven by requests for localized data.

### III. SYSTEM AND SCIENTIFIC APPLICATIONS

*A. Hardware System*

The system we used for the test was the SuperMUC system at the Leibniz Supercomputing Center of the Bavarian Academy of Science (BADW-LRZ). As shown in figure 2 it consists of two independent PFLOP/s machines: SuperMUC phase 1 & SuperMUC phase 2 with a combined peak performance of approximately 6.8PFLOP/s. Both phase 1 and phase 2 systems share a parallel file system, network storage and software environment.

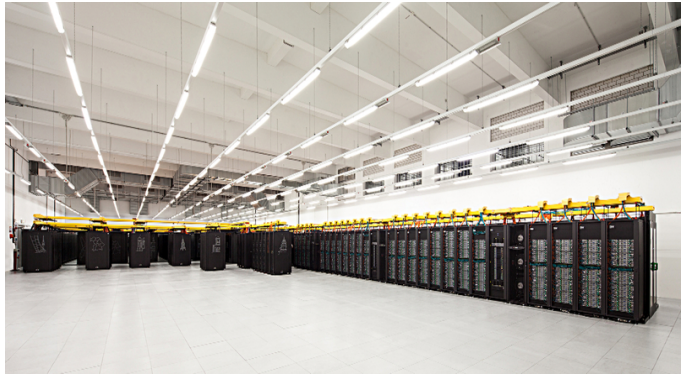

Fig. 2. SuperMUC system at LRZ.

SuperMUC phase 1 is a ~3.2PFLOP/s system consisting of 147,456 Intel Sandy Bridge Xeon E5-2680 8C 2.7GHz cores and 8,200 Intel Westmere Xeon E7-4870 10C cores. The Sandy Bridge cores are arranged into 18 islands consisting of 512 nodes and are referred to as "thin nodes", each node contains 16 cores and each node contains 32GB of memory. While the Westmere cores are arranged in one island consisting of 205 nodes, which are known as "fat nodes" with 40 cores per node and 80 GB of memory, SuperMUC phase 2 consists of 86,016 Intel Haswell Xeon E5-2697v3 cores arranged into 6 islands consisting of 512 nodes, with each node has 28 cores and 64MB of memory.

SuperMUC Phase1 and Phase2 are loosely coupled through the GPFS and NAS File systems, used by both Phase 1 and Phase 2. Both phases are operated independently, but offer an identical programming environment. For high-performance I/O, IBM's General Parallel File System (GPFS) with 12 PB of capacity and an aggregated throughput of 250 GB/s is available.

*B. Scientific Applications*

BQCD (Berlin Quantum Chromodynamics) [13] is a Hybrid Monte-Carlo code written in FORTRAN that simulates QCD with dynamical Wilson fermions and is widely used in the lattice QCD community. The kernel of the program is a standard conjugate gradient solver with even/odd pre-conditioning. In a typical run, between 80% and 95% of the total computing time is used for the multiplication of a very large sparse matrix ("hopping matrix") with a vector. At the single CPU level, QCD programs benefit from the fact that the basic operation is the multiplication of small complex matrices.

SeisSol [14] is a seismic wave propagation solver developed by a research team from the Technical University Munich and the Ludwig-Maximilians-Universität, which is based on the arbitrary high-order accurate Derivative Discontinuous Galerkin (ADER-DG) method on unstructured tetrahedral meshes. It has been successfully applied to simulating various fields of seismology, exploration industry and earthquake physics. SeisSol is able to use all 147,456 cores on SuperMUC phase 1 with a sustained performance of 1.42 PFLOPS, which correlates to 44.5% of theoretical peak performance and a parallel efficiency of 89.7%.

NAMD [15] is a molecular dynamics (MD) program designed for parallel computation. Full and efficient treatment of electrostatic and van der Waals interactions are provided via the ($O$(N Log N)) Particle Mesh Ewald algorithm. NAMD interoperates with CHARMM and X-PLOR as it uses the same force field and includes a rich set of MD features; multiple time stepping, constraints, and dissipative dynamics. NAMD supports MPI, OpenMP and MPI + OpenMP multi-level parallelism, whereas MPI only is the most common parallelism mode on SuperMUC.

GADGET [16] is a hybrid N-body/SPH astrophysics application, which computes gravitational forces with a hierarchical tree algorithm and represents fluids by means of smoothed particle hydrodynamics. As well, as employing an N-body system to simulation gas dynamics.

MGLET [17] has been designed for the numerical simulation of complex turbulent flows. MGLET uses a Finite Volume method to solve the incompressible Navier-Stokes equations. It uses a Cartesian grid with staggered arrangement of the variables that enables an efficient formulation of the spatial approximations. An explicit third order low-storage Runge-Kutta method is used for time integration. The pressure is computed within the framework of fractional step or Chorin's projection method, respectively. Therefore, at every Runge-Kutta substep, a linear system of equations, a Poisson-equation, has to be solved. Geometrically complex surfaces are represented by an Immersed Boundary Method

## IV. RESULTS

In this section, we discuss the analysis of the performance data measurements from the SuperMUC production systems.

### A. BQCD

The performance data measurements were analyzed for two groups, external users and LRZ staff [18], which compared jobs based on the number of CPU used and their execution time. The plot in figure 3 depicts the number of CPU cores and execution times for individual BQCD jobs run on SuperMUC by external users over a two-year period. Notice how the jobs are tightly clustered at 1000 CPU core with an execution time between 100 and 200 seconds.

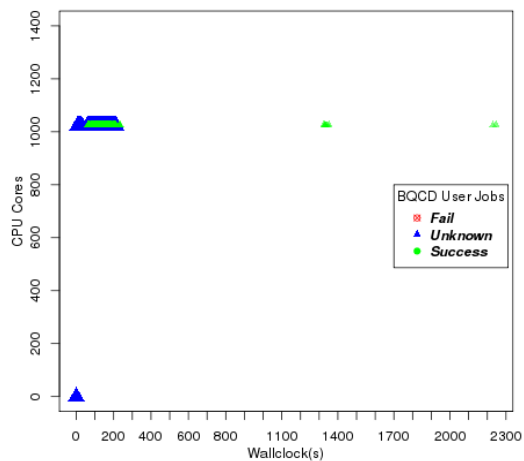

Fig. 3. BQCD jobs for external users.

While LRZ employs BQCD extensively to verify that the system is operating correctly after maintenance, communication software updates and to measure the scaling performance of the system. So the LRZ usage pattern is significantly different to the tightly clustered user jobs, as illustrates in Figure 4.

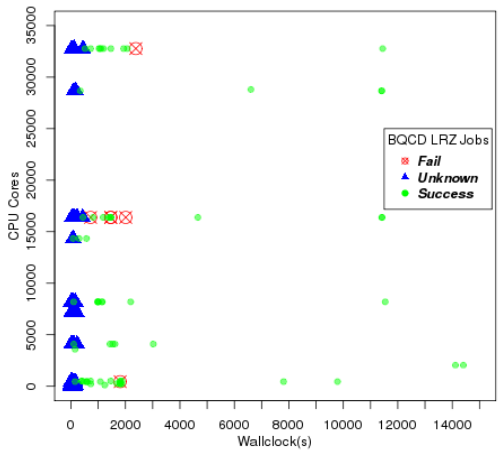

Fig. 4. BQCD LRZ jobs.

To be able to generate enough PerSyst data points, we created a specific BQCD test case that would execute for 4 hours on 128 CPU cores, which produced 25 data points, which we attempted to analyze. In figure 5 we show how the measured FLOP rate in PerSyst varies at the different measurement points. The main numerical algorithim in BQCD is conjugate gradient, the peak FLOP rates correspond to regions of high arithmetic intensity and the low points correspond to regions of communication such as global summation.

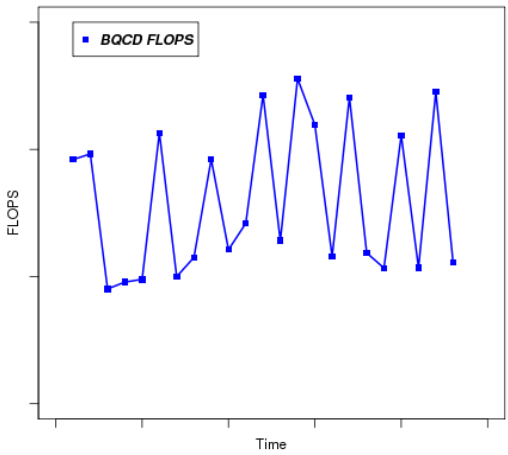

Fig. 5. BQCD branch misprediction to instruction rations ratio.

Figure 6 shows the average cycles per instruction for a 5 second period, which gives an indication of the latency, with higher CPI values meaning there is higher latency. However, vectorization reduces the number of instructions executed overall in your code, but it would likely raise the CPI because the multiple data instructions are more complex and take longer to execute. In many cases, vectorization increases performance, even though CPI went up. As BQCD spends most of its execution time performing matrix multiplication, it is highly vectorized and has good arithmetic performance even though the latency is high.

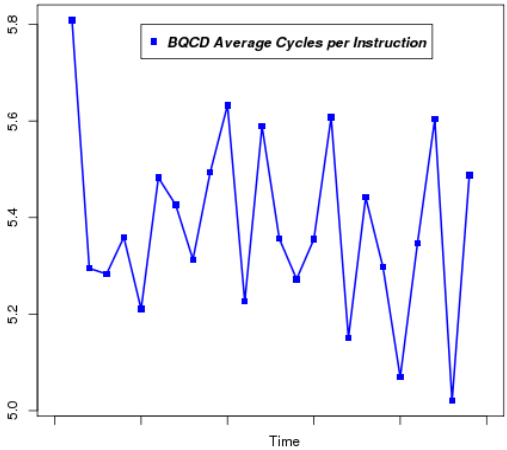

Fig. 6. BQCD average cycles per instruction.

Processors derive much of their performance by executing instructions in parallel and before the time when their results are actually needed. For this to work efficiently with branching code we need to make an accurate prediction of the result of the comparison and executes the instructions that it thinks are next. Figure 7 depicts the ratio of branch mispredictions to instructions during the execution of BQCD, which can possibly be used in conjunction with the cycles per instruction and floating point operations data to determine if an unknown application that has executed on SuperMUC is BQCD. The execution pattern for different BQCD jobs will be very similar as between 80% and 95% of the execution time is spent multiplying large sparse matrices. However, as there are only two successful BQCD jobs that have a long enough execution time to produce enough data, it is difficult to verify this assumption.

Due to the execution time being too short to collect enough data points. BQCD is not a suitable candidate application for the analysis of PerSyst data. As demonstrated by the following execution time statistics: $1^{st}$ quantile (86 S), median (95 S), mean (102 S), $3^{rd}$ quantile (112 S), which would require measurements of two seconds to provide approximately 50 data points.

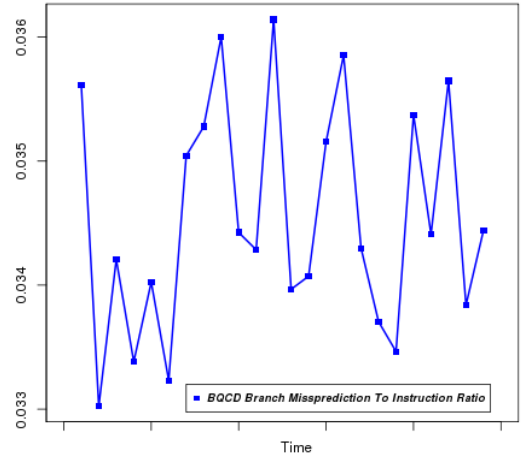

Fig. 7. BQCD branch misprediction to instruction rations ratio.

### B. SeisSol

The performance data measurements were analyzed in a similar way as BQCD for the simulation of the Landers earthquake [19] in 1992 and Sumatra-Andaman Earthquake [20, 21] in 2004.

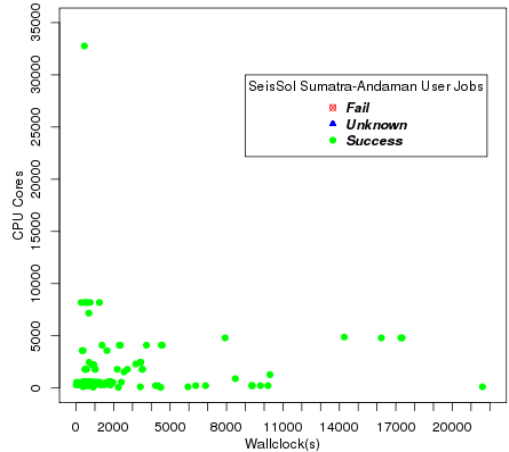

Fig. 8. The SeisSol simulations of the Sumatra-Andaman earthquake.

Figure 8 shows the number of CPU cores and execution times for individual simulations of the Sumatra-Andaman earthquake run on SuperMUC over a two-year period and is a suitable candidate application for the analysis of PerSyst data, due to the average execution time being long enough to collect enough data points. As the demonstrated by the following execution time statistics: 1st quantile (30 S), median (270 S), mean (1184 S), 3rd quantile (1654 S). which means for an average run we would need to take measurements every 20 seconds to produce more than 50 data points.

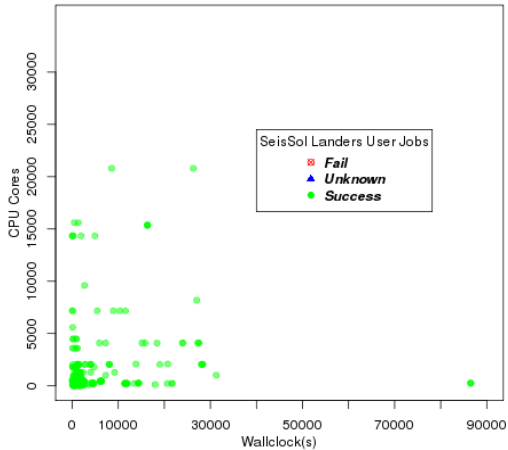

Fig. 9. SeisSol simulation of the Landers earthquake by external.

In a similar way we analyze the 1992 Landers earthquake simulation. The usage pattern of the SeisSol simulation for external users over the two year period depicted in figure 9 shows and clustering at 500 cores and an execution between 200 & 1500 seconds. The execution time statistics for the simulations are: 1st quantile(30 S), median(277 S), mean(1840 S), 3rd quantile(1019 S). So an average simulation of the Landers earthquake would therefore require measurements every 30 seconds to provide more than 50 data points. Figure 10 shows the execution pattern for SeisSol landers simulations done by LRZ. Notice how the runs are not tightly clustered unlike the simulations carried out by external users. This is because LRZ uses SeisSol used for testing purposes and not for generating scientific results.

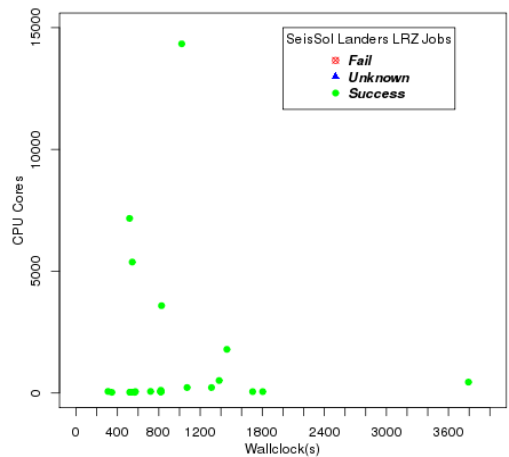

Fig. 10. SeisSol simulation of the Landers earthquake by LRZ.

To examine the data collected by the PerSyst tool we decided to execute the same earthquake simulation on the same version of SeisSol, with one binary built with all the compiler optimization turn on and the other with all compiler optimizations turned off.

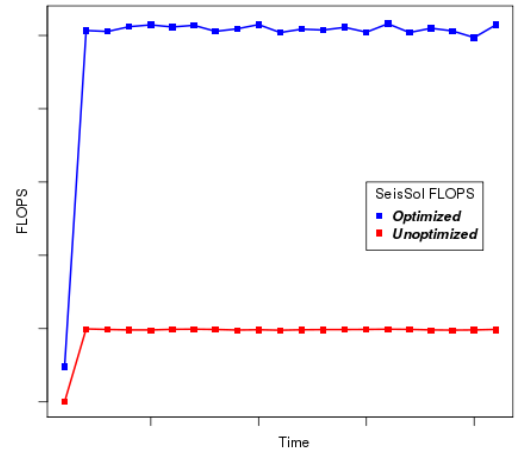

Fig. 11. SeisSol floating point operations per second.

The difference in floating operation performance between the two binaries built with compiler optimizations turned on and off is extremely significant as illustrated in figure 11. This difference is also present in the AVX floating point operation performanceas shown in figure 12 with the unoptimized binary having almost no vectorized instruction. The reason for this discrepancy is that the SeisSol code is highly optimized for the execution of SIMD type instruction, especially for the matrix-matrix multiplication operations, such as the vectorized fused multiply add (FMA) instruction [22, 23]. From other performance measurements of SeisSol, we know that the optimized version has a parallel efficiency of 95% and a 45% peak floating point efficiency on SuperMUC.

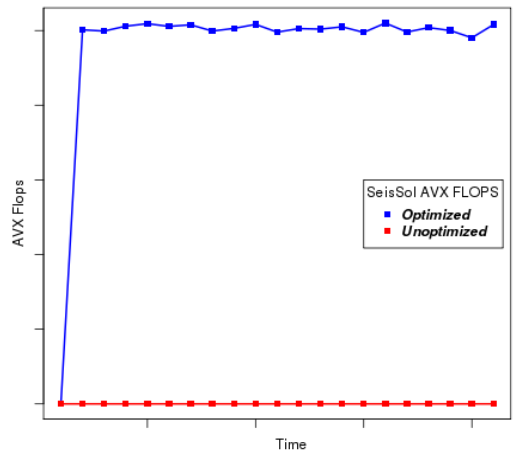

Fig. 12. SeisSol AVX floating points operations per second.

The result of the analysis of the AVX FLOPS measurements stored in the PerSyst database. We are able to determine if a particular SeisSol binary has been built with the correct compiler optimization flags set for SuperMUC.

## C. NAMD

The performance data measurements were analyzed for NAMD in the same way as the previous applications.

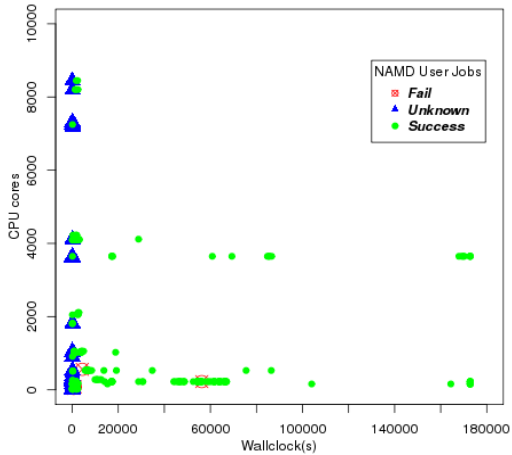

Fig. 13. NAMD simulations by users..

Figure 13 shows how external users execute NAMD and how there are a significant number of jobs that don't produce any PerSyst data, due to the short execution times. Also, there are six jobs whose execution time was sufficient to generate PerSyst data but failed to do so and corresponds to a PerSyst tool failure. Although there appears to be some clustering of NAMD user jobs, it is not as pronounced as for the SeisSol or BQCD jobs. The reason for this is that NAMD has more than one execution mode, so two jobs might be doing very different computation to one another.

LRZ usage of NAMD depicted in figure 14 is typically for testing and verification purposes after updates and system maintenance. That is also the reason why there a relatively more very short execution time jobs, as well as a greater ratio of failures to successful executions compared to external user NAMD jobs.

Due to NAMD being able to calculate different properties and executing different algorithms as well as the median execution time of 95 seconds being too short to collect enough data points, it is not a suitable candidate application for the analysis of PerSyst data.

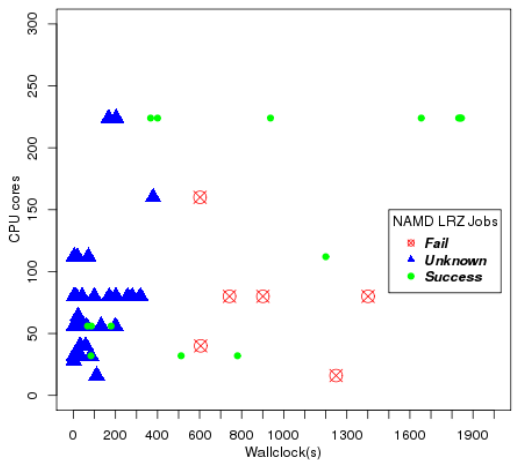

Fig. 14. NAMD simulations by LRZ.

## D. GADGET

The performance data measurements were analyzed for three groups, external users, LRZ staff and an LRZ developer in the same way as the other applications. The manner in which external users utilize GADGET, as illustrated in Figure 15 shows that a significant number of jobs are tightly clustered at between 40 & 600 CPU core; and with an execution time between 20 & 2000 seconds. This result is surprising, as typical cosmological simulations take several hours to complete. One explanation for this is that these particular jobs are for testing modifications to the algorithms. While the jobs that have either a large number of CPUs or run for several hours are what we would typically expect for a complete simulation time step.

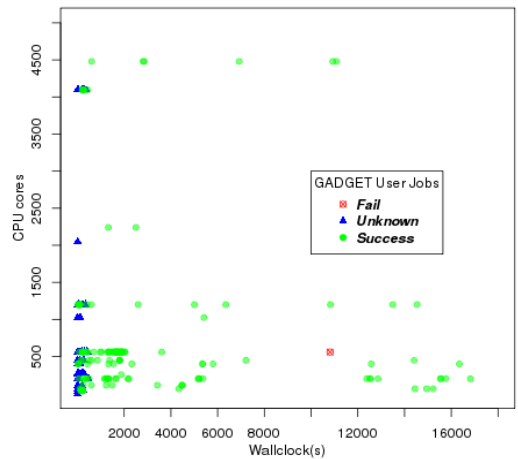

Fig. 15. GADGET simulations by external users.

While LRZ usage of GADGET as shown in Figure 16 is typical for system and basic scaling tests, which is the reason why most of the jobs have execution times less than 5 minutes.

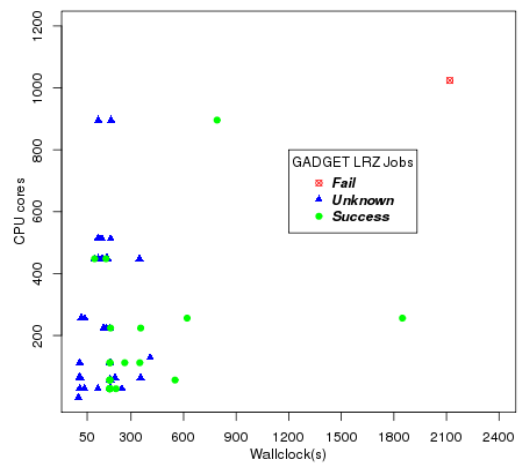

Fig. 16. GADGET simulations by LRZ.

Finally, an LRZ staff member who is working on the development of GADGET utilizes the application in a similar manner to that of the external users and significantly different from how other LRZ users execute GADGET. Notice in Figure 17 that there is significant clustering around the 32-600 CPU cores and 70-2,000 seconds. We also see strong clustering at 16, 256, 640, 1024 and 2048 CPUs, which corresponds to 1, 16, 40, 64 and 128 nodes on SuperMUC. However, there is only a single job with more than 2K CPU cores, while the external users have executed several jobs with 4K CPU cores.

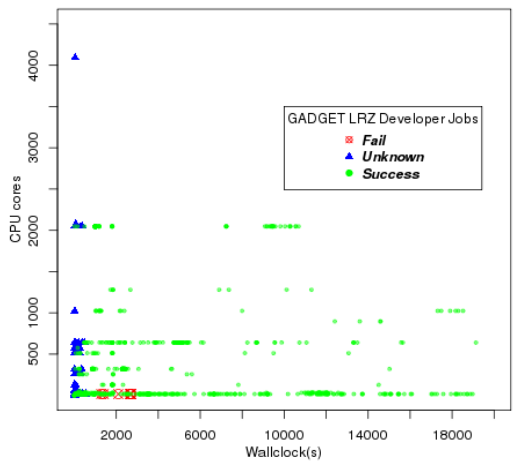

Fig. 17. GADGET simulations by an LRZ developer.

The execution time statistics for simulations whose execution time is greater than 2000 seconds are: $1^{st}$ quantile(5356 S), median(14440 S), mean(36560 S), $3^{rd}$ quantile(54770 S). So an average GADGET simulation would generate 61 data points at a sampling rate of every 10 minutes.

*E. MGLET*

The performance data measurements of MGLET were analyzed for three groups; LRZ internal useage and two different external user projects. Figure 18 depicts how LRZ utilize MGLET and we can see that a significant number of jobs are clustered at 16, 28 and 112 CPU cores, which corresponds to 1 node on SuperMUC phase 1 and 1 & 4 nodes on SuperMUC phase 2.

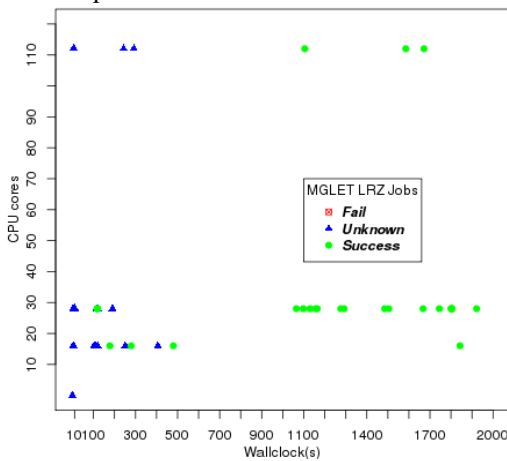

Fig. 18. MGLET simulations by LRZ.

Figure 19 illustrates how the project pr63zi executes on SuperMUC phase 1. Notice how there is strong clustering at 128 CPU cores, which corresponds to 8 nodes on SuperMUC phase 1. As well as some clustering at 66 seconds, which suggests that this project only runs small CFD simulations.

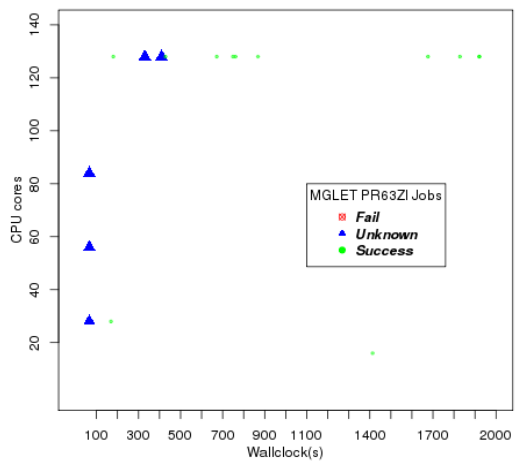

Fig. 19. MGLET simulations by project pr63zi.

While the project pr84gi as illustrated in figure 20 runs a wide range of different sized CFD simulations. Notice that there is a large clustering of jobs at 512 and 1024 CPU cores. Also, there is a large number of jobs that have an execution time of less than 10,000 seconds and has less than 5,000 CPU cores.

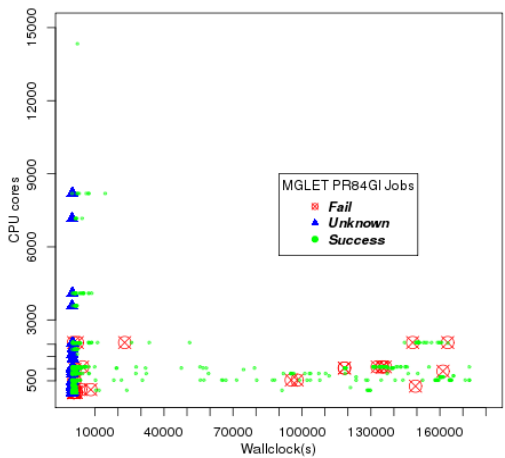

Fig. 20. MGLET simulations by project pr84gi.

Finally, we look at the jobs that use less than 4,500 CPU cores and have an execution time of less than 5,000 seconds by the project pr84gi. We can see strong clustering at 16, 28, 128, 224, 512, 528, 800, 1024 and 2048 CPU cores. As well as strong clustering of the jobs whose execution time is below 2000 seconds. If we compare how the two different user projects execute MGLET we notice that they are significantly different and that pr84gi has a significant spread of jobs in both number of CPU cored used and the applications execution time. While pr63zi has a much lower number of jobs run and whose number of CPU cores and execution are much more tightly bunched.

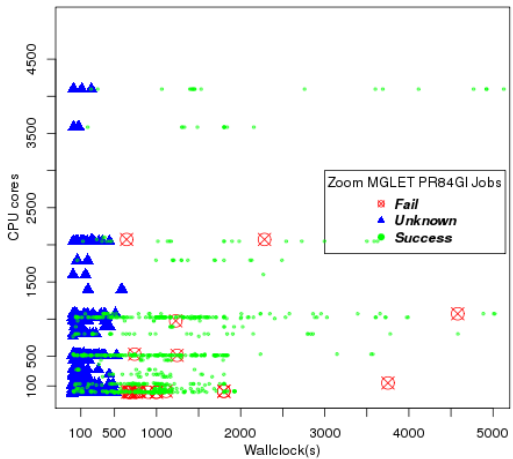

Fig. 21. Subset of MGLET simulations by project pr84gi.

V. CONCLUSIONS

The analysis of application performance data on HPC systems is very important in the allocation of resources and whether the system is being used as intended. We have shown that collecting performance data every 10 minutes does not provide enough information to allow advanced data analytics to be employed, because for a significant number of application runs we only collect a few data points. To overcome this issue the system monitoring tool would have to be configured to collect data at least every 30 seconds, which might not be possible on a production system due to system performance and data storage overheads. However, we are able to use the data collected to determine how the scientific applications are executed on the system and which applications would be suitable candidates for future performance predictions.

FUTURE WORK

We would like to extend this work to identifying improvements to system utilization and resource management. As well as using machine learning techniques on the scientific applications that have been identified as suitable candidates.

ACKNOWLEDGMENT

The authors would like to thank The Leibniz Supercomputing Centre Bavarian Academy of Sciences and Humanities for granting us access to a petaflop system.

Kunkel, P. Balaji, and J. Dongarra, Editors. 2016, Springer International Publishing: Cham. p. 343-362.